\documentclass[reprint,amsmath,amssymb,aps,pre]{revtex4-1}
\usepackage{graphicx} 
\usepackage{dcolumn}   
\usepackage{bm}        
\usepackage{amssymb}   
\usepackage{mathtools}
\usepackage{amsmath}
\usepackage{color}
\usepackage{dblfloatfix} 
\usepackage{hyperref}
\hypersetup{
    citecolor={cyan},
    colorlinks=true,
    linkcolor=cyan,
    filecolor=magenta,      
    urlcolor=cyan,
    pdfpagemode=FullScreen,
    }
\begin{document}

\title{Rectification of Random Walkers Induced by Energy Flow at Boundaries}
\author{Vidyesh Rao Anisetti$^1$, Sharath Ananthamurthy$^2$, J. M. Schwarz$^{1,3}$}

\affiliation{$^1$ Physics Department, Syracuse University, Syracuse, NY 13244 USA
\\$^2$ School of Physics, University of Hyderabad, Hyderabad 500046, India \\
$^3$ Indian Creek Farm, Ithaca, NY 14850 USA }
\begin{abstract}
    We explore rectification phenomena in a system where two-dimensional random walkers interact with a funnel-shaped ratchet under two distinct classes of reflection rules. The two classes include the angle of reflection exceeding the angle of incidence ($\theta_{reflect} > \theta_{incident}$), or vice versa ($\theta_{reflect} < \theta_{incident}$). These generalized boundary reflection rules are indicative of non-equilibrium conditions due to the introduction of energy flows at the boundary. Our findings reveal that the nature of such particle-wall interactions dictates the system's behavior: the funnel either acts as a pump, directing flow, or as a collector, demonstrating a ratchet reversal. Importantly, we provide a geometric proof elucidating the underlying mechanism of rectification, thereby offering insights into why certain interactions lead to directed motion, while others do not.  
\end{abstract}

\maketitle

\section{Introduction}
Systems in thermal equilibrium do not show rectification as demonstrated by the Feynman–Smoluchowski ratchet\cite{feynman1963feynman,smoluchowski1912experimentell}. However, non-equilibrium systems with an underlying spatial asymmetry do exhibit sustained motion rectification, or motor-like behavior~\cite{hanggi2009artificial,feynman1963feynman}.
An experimental realization of what is now known as a Brownian motor consists of four vanes that can freely rotate and surrounded by a vibrated granular gas~\cite{eshuis2010experimental}. This system is out-of-equilibrium as the collisions are inelastic and granular particles are driven by external vibrations (and not thermal fluctuations). Of course, this example naturally leads one to the field of active matter, in which each system constituent consumes energy to lead to self-generated motion~\cite{ramaswamy2010mechanics,marchetti2013hydrodynamics}. Indeed, creating molecular motors/engines using active Brownian particles, be it living or nonliving, has been explored by many~\cite{costantini2007granular,Tailleur2009,Sekimoto1998}. One such experimental example consists of active bacterial baths being used to operate asymmetric gears~\cite{Sokolov2010,Leonardo}.

One of the simplest forms of an active engine is motion rectification of active matter in the presence of funnel-shaped ratchets~\cite{galajda,wan}.  Interestingly, it has been shown that active, or self-propelled, particles show rectification in presence of funnel-shaped ratchets because of breaking of detailed balance which occurs when particles slide along the boundary after encountering it~\cite{wan}. However, for other types of particle-boundary interactions, for example, pure reflection, rectification is lost, despite the particles being active~\cite{Tailleur2009,10.1117/12.897424}. Therefore, we go back to a simpler system of non-active Brownian particles and ask the question: What types of particle-boundary interaction rules lead to motion rectification? 

\begin{figure}[h!]
    \centering
    \includegraphics[width=8cm]{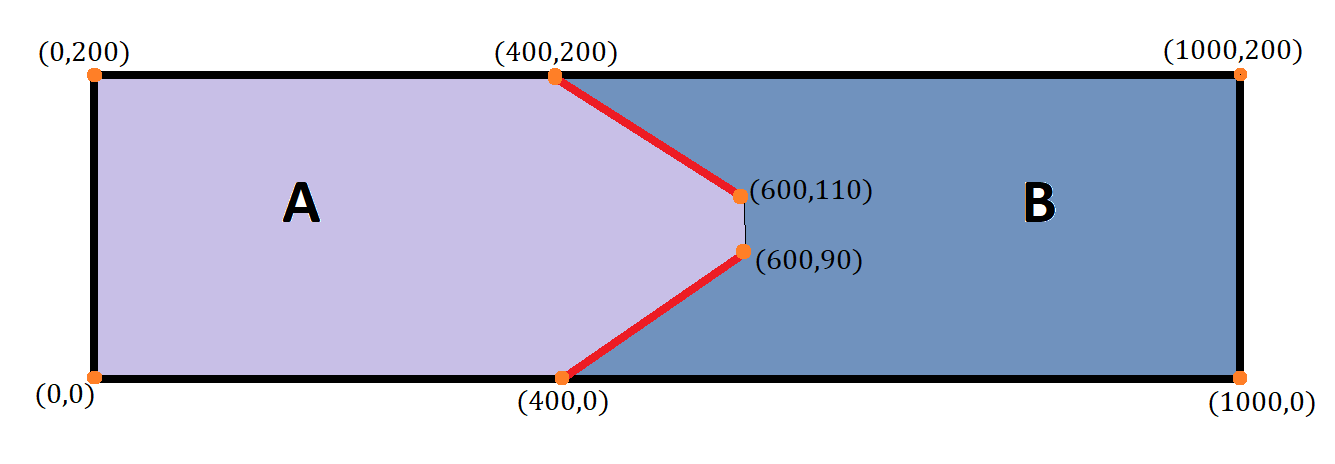}
    \caption{The shaded areas specify region A and B respectively. Coordinates of orange dots are shown.
    }   
    \label{fig:dimensions}
\end{figure}

By investigating this question, we show that there exists a class of particle-boundary interactions that gives rise to rectification, and sliding along the boundary is simply a special case of this class. Additionally, we provide a physical understanding behind this effect, by using particle kinematics and geometry of the boundary we provide a geometric proof on why, rectification occurs for this class of interaction and not for pure reflection. Our approach simplifies the system to its core processes by emphasizing only the essential properties responsible for rectification. This primary simplification entails simulating particle kinematics independently of the forces guiding their trajectories, which is a departure from conventional methods~\cite{wan}. Our model system consists of a two-dimensional rectangular chamber with a single funnel-shaped ratchet in between (Fig. \ref{fig:dimensions}). Within this chamber, we introduce non-interacting random walkers. While these walkers obey the reflection rule upon contacting the rectangular boundary, they exhibit modified reflection behavior when interacting with the funnel. 

We define two classes of this modified reflection:-
\begin{align}
     & (i)~  \theta_{r}=\theta_{i} + \alpha\left(\dfrac{\pi}{2}-\theta_{i} \right),\\ 
     & (ii)~ \theta_{r}=\theta_{i} - \alpha\theta_{i} ~; \alpha \in [0,1] 
\end{align}
Here, $\theta_r$ and $\theta_i$ denote the angles of reflection and incidence, respectively. Rule (i) results in $\theta_{r} > \theta_{i}$ while Rule (ii) leads to $\theta_{r} < \theta_{i}$. Each value of $\alpha$ corresponds to a specific reflection rule. The parameter $\alpha$ modulates the extent of deviation from the standard law of reflection. When $\alpha = 0$, the modified rule reverts to $\theta_{r}=\theta_{i}$. Notably, for $\alpha=1$ in Rule (i), the condition simulates particle sliding post-collision with the funnel, a scenario explored in prior research with active particles~\cite{wan, 10.1117/12.897424, Tailleur2009}. 

The rest of the manuscript is organized as follows.  We detail the simulation methodology, then present our simulation results. A simple, geometric proof helps to interpret our simulation results. We conclude with a discussion of the implications of our findings. 
\begin{figure}[h!]
    \centering
     \label{fig:rule}\includegraphics[width=9cm]{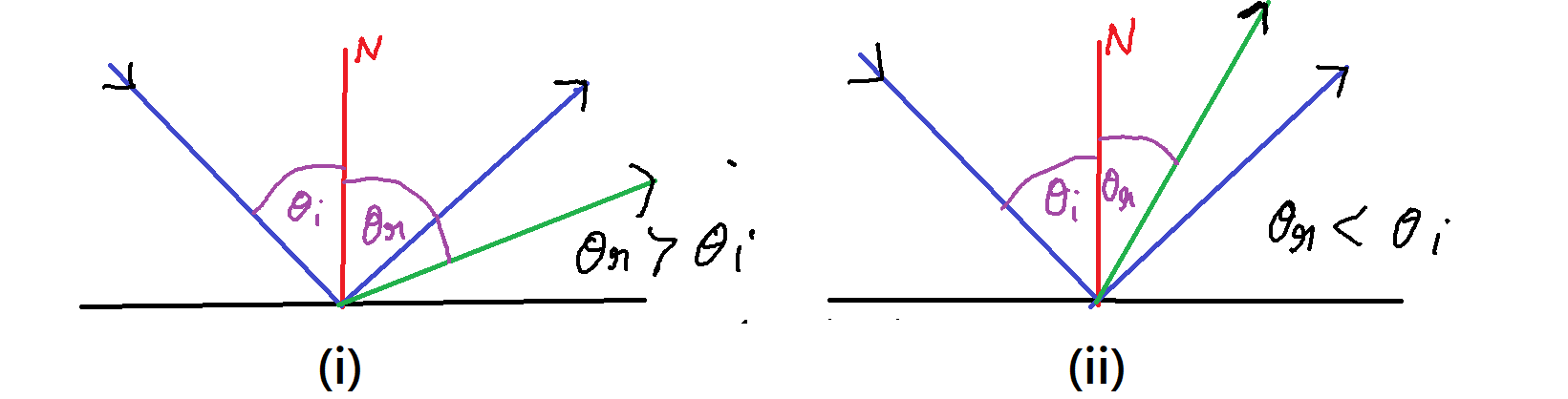}
    \caption{The left image shows an example of Rule (i) and right image shows an example of Rule (ii). The blue outgoing arrow shows the path the particle would have followed in case of perfect reflection rule. The green outgoing arrow shows the path which particle takes in case of modified reflection rule. Notice that in Rule (i) the particle is deviated towards the wall and in Rule (ii) the particle is deviated towards the normal.
    }   
    \label{fig:1}
\end{figure}
\section{Simulations}
The system consists of a rectangular box of dimensions 1000 $ \times $ 200, with a single funnel in the middle (Fig. \ref{fig:dimensions}). This is different from the geometry studied in Ref. \cite{wan} \cite{galajda}, which used multiple funnels. In this system, we study 5000 non-interacting random walkers following the iterative equation :- 
\begin{align}
   & x_{t+dt}=x_t + \lambda \sin(2\pi \zeta(t)) \\
   & y_{t+dt}=y_t + \lambda \cos(2\pi \zeta(t)),
\end{align}
where $(x_t,y_t)$ define the position of each particle at time $t$ and $\zeta$ is a function that outputs a random number  $\zeta(t) \in [0,1]$ with a uniform distribution. Moreover, $\lambda$ is the persistence length of these random walkers. In the simulations, $dt=1$. If a particle encounters the rectangular boundary during a time step, it reflects off it according to the standard law of reflection. However, upon interacting with the funnel, it follows one of the modified reflection rules detailed earlier. If a particle, transitioning from $(x_t,y_t)$ to $(x_{t+1},y_{t+1})$, intersects the funnel before completing its step, the point $(x_{t+1},y_{t+1})$ is reflected about the funnel. This point is then rotated about the collision point based on the reflection rule and $\alpha$ value, as shown in Fig. \ref{fig:rule}. It is crucial to note that each particle traverses a distance equal to $\lambda$ in each time step, even during a collision.

During the simulations, we observed that certain particles tend to collide asymptotically with the corners in region B, where the funnel merges with the rectangle. These particles generally follow Rule (i) with a high $\alpha$ value. When such a particle approaches a corner within a distance of less than 0.1 units, it receives a directional ``kick". This kick ensures that the particle covers a distance $\lambda$ in one time step, while also ensuring that the particle's direction lies within the apical angle of the corner.
\section{Results}
\begin{figure}[h!]
    \centering
    \includegraphics[width=9cm]{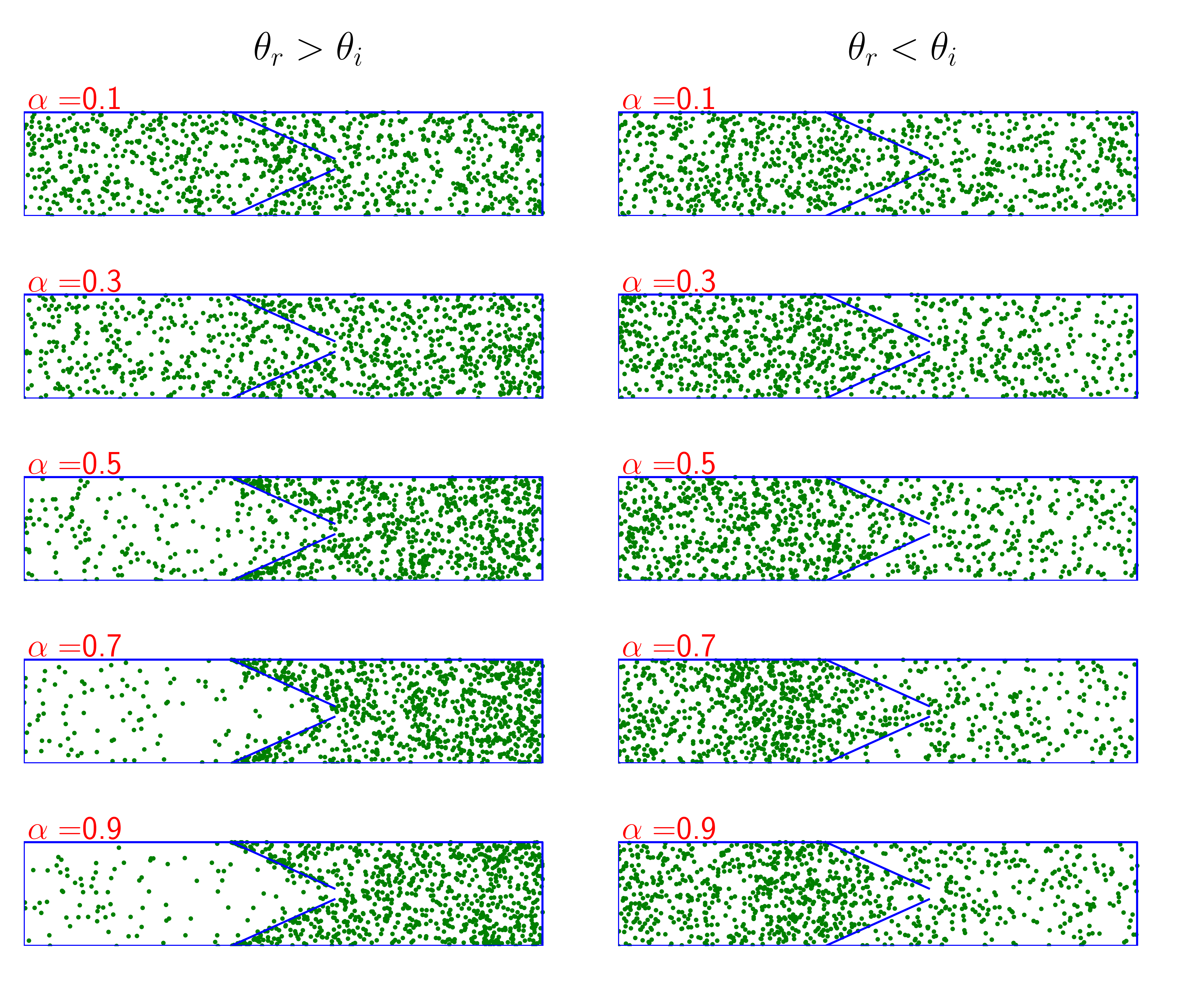}
    \caption{Snapshot of the simulation at t=1000 for for rules (i) (left column)  (ii) (right column) respectively for different $\alpha$s.}   
    \label{fig:snapshot}
\end{figure}

The system is initialized with a uniform number density of random walkers throughout the system. Gradually, we see number density increase in one of the chambers depending on the reflection rule (see Fig.~\ref{fig:snapshot}). We observe that an increase in $\alpha$ leads to a higher left-right number density asymmetry in the system at steady state.  Additionally, within each chamber, a higher $\alpha$ value results in a more heterogeneous steady-state number density distribution. For $\alpha=0$, signifying perfect reflection, the particles exhibit a uniform distribution across the system. The particles in the system are initialized with a uniform number density.

\begin{figure}[h!]
    \centering
    \includegraphics[width=9cm]{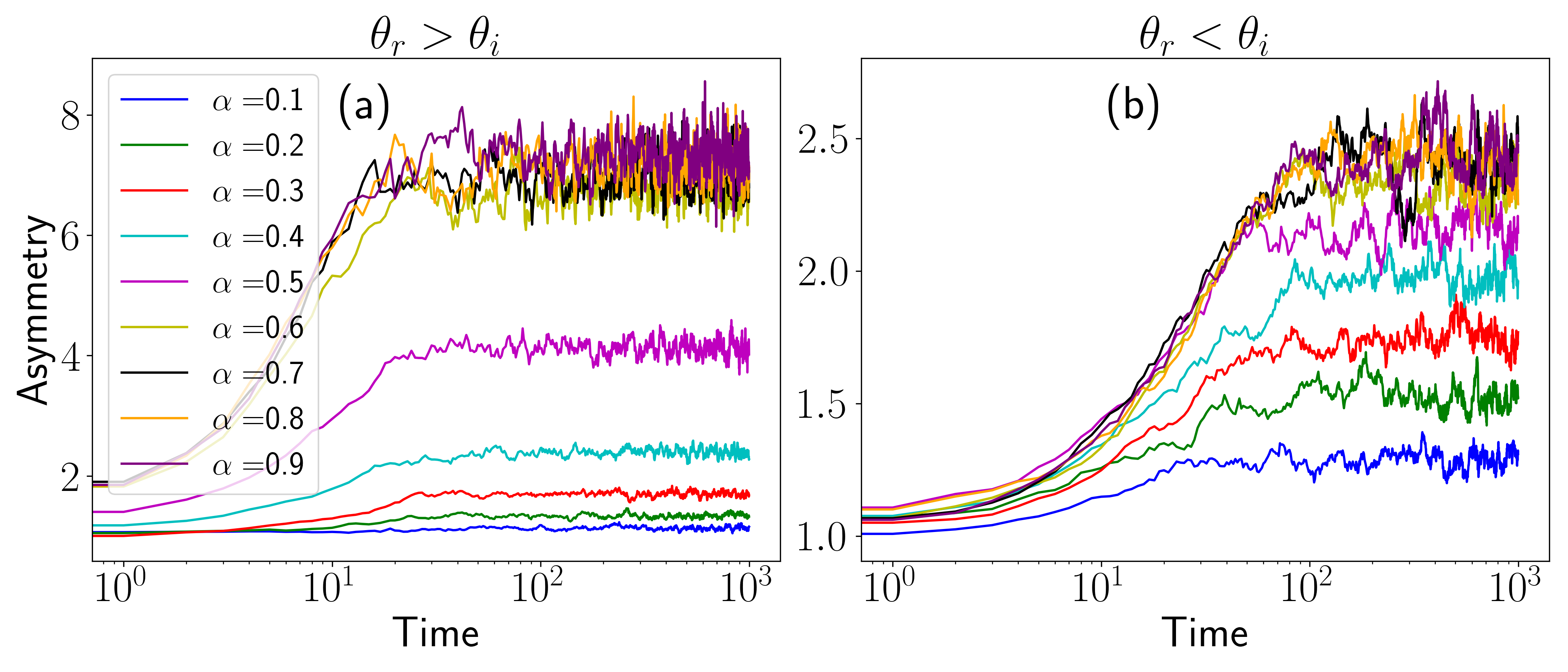}
    \caption{Plots shows asymmetry defined as $\dfrac{max(n_a,n_b)}{min(n_a,n_b)}$ with time for different $\alpha$ where plot (a) and (b) show 
Rule (i) and (ii) respectively.}    \label{fig:ratio-vs-time}
\end{figure}

For Rule (i), we observe a higher particle accumulation in chamber B, with the number density ratio escalating with increasing $\alpha$. This suggests that the funnel acts as a pump, directing particles into chamber B (see Fig. \ref{fig:snapshot}). Conversely, for Rule (ii), more particles gather in chamber A. This indicates a net movement of particles against the funnel's ``easy" direction, signifying a ratchet reversal. We further analyzed how the ratio of particle number density across the funnel evolves over time (Fig. \ref{fig:ratio-vs-time}). For the same $\alpha$ value, Rule (i) induces a greater asymmetry at steady state than Rule (ii).

To understand this phenomenon fundamentally we need to understand why there is no rectification in case of pure reflection. In the next section we provide a geometric proof of this. 
\begin{figure}[h]
    \centering
    \includegraphics[width=9cm]{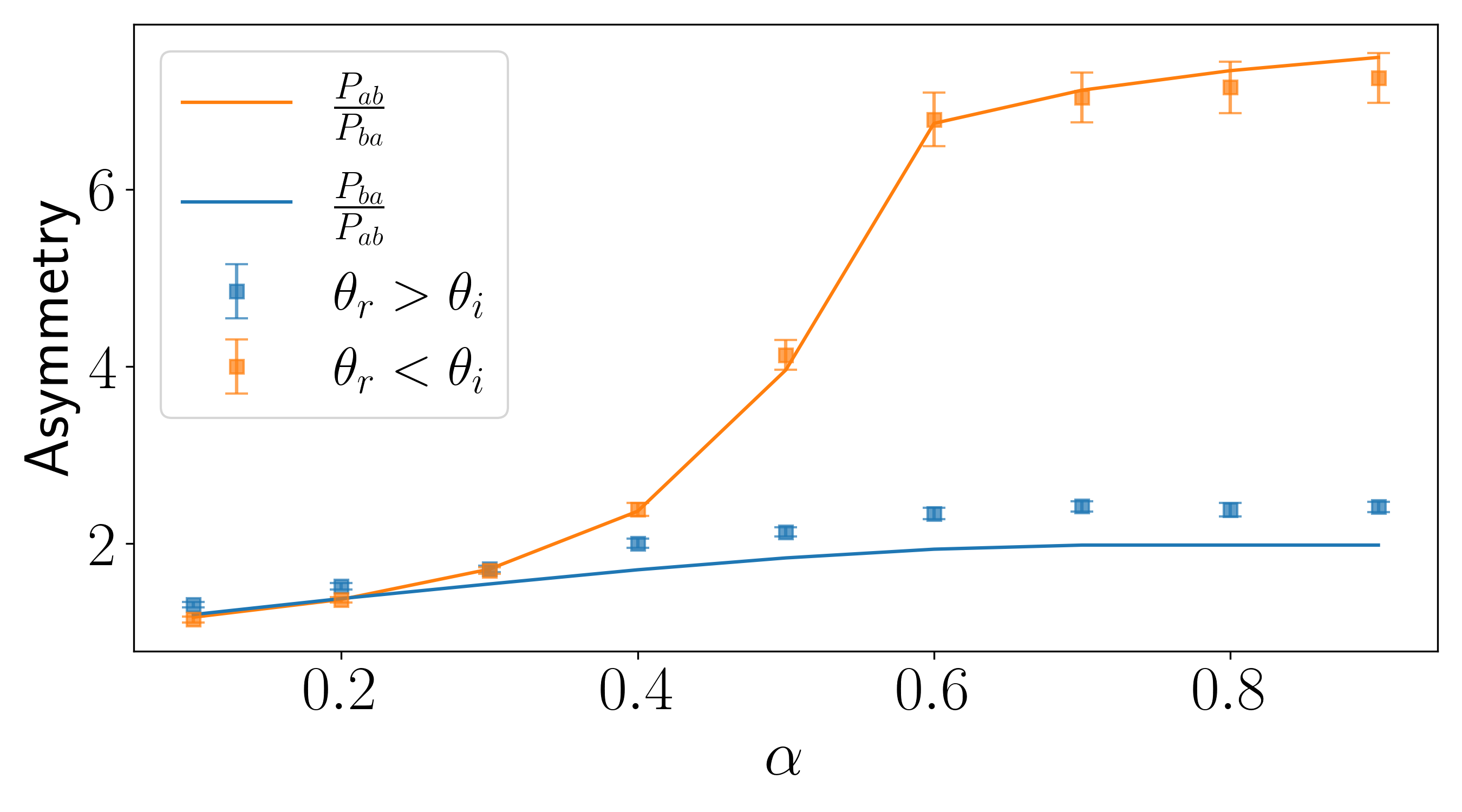}
    \caption{The dotted lines show the steady state number density ratio found from simulation and the solid line is $ \dfrac{ P_{A\xrightarrow{}B}}{ P_{B\xrightarrow{}A}}$.  }
    \label{fig:simul-theor-compare}
\end{figure}
\begin{figure*}[!ht]
    \centering   
    \includegraphics[width=13cm]{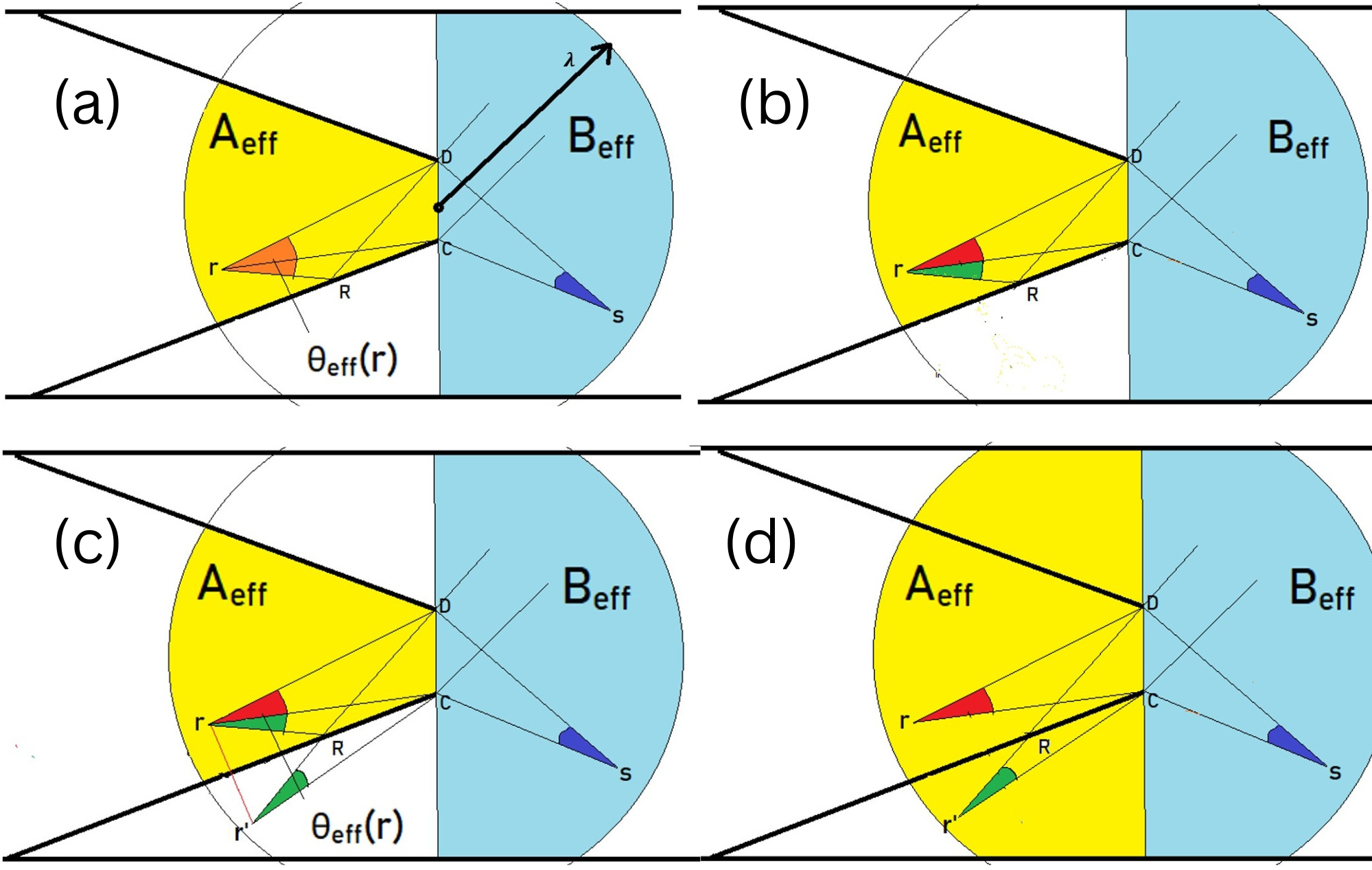}
    \caption{\textbf{Understanding the Absence of Rectification in Purely Reflecting Particles:}\textbf{(a)}In region A,  particle at $\Vec{r}$ can transition directly into region B in one step, or hit the funnel between points R and C to subsequently enter region B.  This range of angles over which this transfer can happen is $\theta_{eff}$ ( shaded in orange ). Conversely, a point s in region B exhibits a smaller range of directions (depicted in dark blue). The effective areas, $A_{eff}$ (in yellow) and $B_{eff}$  (in light blue), demarcate the regions where $p(\vec{r})$ is non-zero in regions A and B, respectively. It's important to note that
     $B_{eff}$ is larger than $A_{eff}$, showing that the funnel has restricted some particles that could go in region B. \textbf{ (b)} : For a point $\Vec{r} \in A_{eff}$, the effective angle, $\theta_{eff}(\Vec{r})$, is the sum of $\theta(\Vec{r})$ (in red) and $\theta_{ref}(\Vec{r})$ (in green).\textbf{ (c)} :
     Consider a point $\Vec{r}^{~\prime}$, which is reflection of point $\Vec{r}$ about the funnel's lower arm. This point subtends an angle, $\theta(\Vec{r'})$, with respect to opening CD. As we are dealing with the case of pure reflection, this angle corresponds to , $\theta(\Vec{r}^{~\prime}) = \theta_{ref}(\Vec{r})$. This implies that a particle originating from $\Vec{r}$ and reflecting off RC to pass through CD has the same probability as a hypothetical particle starting at $\Vec{r}^{~\prime}$ and moving through CD, ignoring the funnel's presence. This reasoning is extendable to cases of multiple reflections. \textbf{ (d)} : So now instead of adding $\theta_{ref}(\Vec{r})$ to account for increase in probability we can ignore the funnel and rather extend the domain of integration to include the points beyond the funnel. Consequently, $A_{eff}$ and $B_{eff}$ are symmetrical with respect to the line through CD. This symmetry ensures that the integrals $P_{A\xrightarrow{}B}$ and $P_{B\xrightarrow{}A}$ (referenced in Eq.~\ref{eq:transfer-probabilities}) are equivalent.}
    \label{fig:geometric_proof}
\end{figure*}

\section{A Geometric Proof}
\subsection{Why $\alpha=0$ shows no rectification}
For pure reflection, where $\alpha=0$, we evaluate the one-step transfer probabilities, $P_{A\xrightarrow{}B}$ and $P_{B\xrightarrow{}A}$, of particles transitioning between chambers A and B. Through a simple geometric proof, we show why purely reflecting random walkers do not show rectification in the presence of asymmetric boundaries. In the preceding subsection we explain why we see rectification along the easy direction of funnel in Rule (i) and a ratchet reversal in Rule (ii).

The probability of a particle at $\vec{r}$ transitioning to the adjacent chamber in a single step is:
\begin{equation}
    p(\Vec{r}) = 
     \begin{cases}
\text{$\dfrac{\theta_{\text{eff}}(\vec{r})}{2\pi}  $} & \text{if } \lambda \ge R \\
       \text{0} & \text{if } \lambda < R,\\
     \end{cases}
\end{equation}
where $R$ represents the distance between $\Vec{r}$ and the nearest point on the opening, and $\theta_{\text{eff}}$ designates the angular range facilitating chamber transition. This effective angle, $\theta_{\text{eff}}(\Vec{r})$, comprises contributions from both direct and reflective transfers, as visualized in Fig.~\ref{fig:geometric_proof}(a). The one-step transfer probabilities are then:\begin{equation}
    P_{A\xrightarrow{}B} = \int_A \dfrac{\theta_{eff}(\vec{r})}{2\pi} \,d^{2}r~~~;~~~
    P_{B\xrightarrow{}A}= \int_B \dfrac{\theta_{eff}(\vec{r})}{2\pi} \,d^{2}r
    \label{eq:transfer-probabilities}
\end{equation}

Upon examining Fig.~\ref{fig:geometric_proof}(a), it looks like the transfer probabilities $P_{A\xrightarrow{}B}$ and $P_{B\xrightarrow{}A}$ aren't inherently equal. In region A, the value of $p(\vec{r})$ is elevated because reflections expand the angular range, augmenting particle transitions to the adjacent chamber. In contrast, region B has a larger area with non-zero $p(\vec{r})$. Intriguingly, the enhanced probability in $P_{A\xrightarrow{}B}$ (attributable to reflection) is precisely offset by the increase in $P_{B\xrightarrow{}A}$ owing to the larger effective area $B_{\text{eff}}$ (Fig.~\ref{fig:geometric_proof}(c-d)) .\\ \\
This observation underscores a pivotal point: irrespective of the funnel's geometry, regions A and B will maintain equal particle number densities. This is solely possible due to perfect reflection.

\subsection{Why $\alpha \neq 0$  is necessary for rectification}
To understand what happens in case of asymmetric reflection we have to go back to Fig.~\ref{fig:geometric_proof}(c). For Rule (i), $\theta_{r}>\theta_{i}$, this means now the particles with smaller incident angles can go into region B, this will result in larger $\theta_{ref}(\vec{r})$ as particles hitting the funnel to the left of point R can now enter region B. Therefore now $\theta_{ref}(\vec{r})>\theta(\vec{r'})$ which leads to $P_{A\xrightarrow{}B}>P_{B\xrightarrow{}A}$. On the flip side, for Rule (ii), we get $\theta_{ref}(\vec{r})<\theta(\vec{r'})$ which leads to $P_{A\xrightarrow{}B}<P_{B\xrightarrow{}A}$.

To validate these deductions, we conducted numerical calculations for $P_{A\xrightarrow{}B}$ and $P_{B\xrightarrow{}A}$, juxtaposing them with the steady-state number density ratios derived from our simulations, as depicted in Fig.~\ref{fig:simul-theor-compare}. In equilibrium, particles traversing the funnel from either side are balanced, which translates to the relation $n_{A}P_{A\xrightarrow{}B} = n_{B}P_{B\xrightarrow{}A}$. While our observations align well for $\alpha>0$, discrepancies arise for $\alpha<0$. This divergence might stem from the prolonged time systems with $\alpha<0$ require to reach a steady state, as shown in Fig.~\ref{fig:ratio-vs-time}. Hence, using one-step probabilities might not effectively capture the number density ratios in these cases.

Delving deeper, when $\alpha>0$, particles experience a force directing them towards the boundary, leading to accumulation of particles in the corners of the box. Conversely, for $\alpha<0$, particles are repelled from the boundary, moving away from the corners, an effect visible in Fig.~\ref{fig:snapshot}. The scenario for $\alpha=0$ is distinct, as the density distribution of purely reflecting particles remains uniform, irrespective of boundary geometry.

\section{Discussion}
Purely reflecting random walkers do not show rectification even for funnel-shaped boundary. To observe rectification, we introduce a non-equilibrium effect in the system by modifying the reflection rules. When the boundary reflection rule deviates from pure reflection, time-reversal symmetry is broken, particle-boundary interactions become non-reciprocal, which results in the ratchet acting as a Maxwell demon.

Our result helps connects the different rectification phenomenon that has been observed in self-propelled particles, ballistic chains,  flexible vesicles , granular gases, and even thermal systems with nontrivial interactions~\cite{galajda,wan2013directed,gandikota2023rectification,solorzano2017thermal,costantini2007granular}.  Self- propelled particles that slide along the boundary after collision follow Rule (i) with $\alpha=1$~\cite{galajda}, and therefore show rectification along the easy direction of funnel. Ballistic chains when interacting elastically with the boundary show an effective reflection law that looks like Rule (ii), which explains why a ratchet reversal was observed~\cite{wan2013directed}. Deviations from pure reflection can be experimentally achieved by introducing a temperature difference between the gas molecules and the collision surface~\cite{molecular_beam1966}, resulting in an exchange of energy during collision and making the particle-wall interaction ``active''.  Such experimental realizations have the potential to pave the way for innovative engines that harness these non-equilibrium phenomena, potentially leading to the creation of highly efficient engines reminiscent of molecular motors.

\bibliography{biblio.bib} 
\bibliographystyle{ieeetr}
\end{document}